\documentstyle[epsfig]{mn}
\newif\ifAMStwofonts
\AMStwofontstrue
\def\be{\begin{equation}}
\def\ee{\end{equation}}

\def\gtsima{$\; \buildrel > \over \sim \;$}
\def\ltsima{$\; \buildrel < \over \sim \;$}
\def\prosima{$\; \buildrel \propto \over \sim \;$}
\def\gsim{\lower.5ex\hbox{\gtsima}}
\def\lsim{\lower.5ex\hbox{\ltsima}}
\def\simgt{\lower.5ex\hbox{\gtsima}}
\def\simlt{\lower.5ex\hbox{\ltsima}}
\def\simpr{\lower.5ex\hbox{\prosima}}

\def\ie{{\frenchspacing\it i.e. }}

\title[Constraining DM through 21 cm observations]{Constraining dark matter through 21 cm observations}         
\author[Vald\'{e}s, Ferrara, Mapelli, Ripamonti]
{M. Vald\'{e}s$^1$, A. Ferrara$^1$, M. Mapelli$^1$, E. Ripamonti$^2$\\
$^1$ SISSA/ISAS, via Beirut 2-4, 34014 Trieste, Italy;\\
$^2$Kapteyn Astronomical Institute, University of Groningen, Postbus
800, 9700 AV, Groningen, The Netherlands.}

\begin{document}
\maketitle
\label{firstpage}
\begin{abstract}
Beyond reionization epoch cosmic hydrogen is neutral and can be directly observed through its 21~cm
line signal. If dark matter (DM) decays or annihilates the corresponding energy input affects the hydrogen 
kinetic temperature and ionized fraction, and contributes to the Ly$\alpha$ background.
The changes induced by these processes on the 21~cm signal can then be used to constrain 
the proposed DM candidates, among which we select the three most popular ones:
(i) 25-keV decaying sterile neutrinos, (ii) 10-MeV decaying light dark matter (LDM) and (iii) 10-MeV annihilating LDM. 
Although we find that the DM effects are considerably smaller than found by previous studies (due to a more
physical description of the energy transfer from DM to the gas), we conclude that 
combined observations of the 21~cm background and of its gradient should be able to put constrains at least on
LDM candidates. In fact, LDM decays (annihilations) induce differential brightness temperature variations 
with respect to the non decaying/annihilating DM case up to $\Delta \delta T_b=8 \,\, (22)$~mK at about 50 (15)~MHz.
In principle this signal could be detected both by current single dish radio telescopes and future
facilities as LOFAR; however, this assumes that ionospheric, interference and 
foreground issues can be properly taken care of.    
\end{abstract}


\section{Introduction}

Cosmic recombination at $z\sim $ 1000 left the gas in the Universe in a nearly uniform, dark, neutral state
which we currently denote as the Dark Ages of the Universe. Gas remained neutral until the first luminous 
objects emerged at $z\sim $ 20$-$30, leading to the complete reionization of the intergalactic medium (IGM) 
at a later, yet unknown, epoch.  

The investigation of the Dark Ages is one of the frontiers of modern cosmology and a new generation of 
radio interferometers is currently in development to study the redshifted 21~cm hyperfine triplet-singlet 
level transition of the ground state of neutral hydrogen. Instruments such as 
the LOw Frequency ARray (LOFAR), the 21 Centimeter Array (21CMA), the Mileura Wide-field 
Array (MWA) and the Square Kilometer Array (SKA), are expected to reach the sensitivity required to map the HI 
distribution at angular resolution of the order of a few arcminutes (e.g. Pen, Wu, \& Peterson 2004; Bowman, Morales, 
\& Hewitt 2005; Kassim et al. 2004; Wyithe, Loeb, \& Barnes 2005).

The main scientific goal of such instruments is to perform an accurate tomography of matter during these remote 
cosmic epochs, and reconstruct in detail the latest phases of the reionization history. However, this might not
be the only major achievement of such experiments.  For example, an analysis of the features of HI 21~cm maps from 
the Dark Ages could also allow to put constrains on the existence and nature of decaying and annihilating Dark 
Matter (DM).  In fact, if DM actually decays or annihilates, the injection of high energy photons into the IGM would 
heat and ionize the neutral gas, leaving an imprint on the 21~cm background signal which could be directly observed
(Shchekinov \& Vasiliev 2006; Furlanetto, Oh, \& Pierpaoli 2006).

The importance of DM decays and annihilations on the reionization history has been analyzed in detail by several
authors. Most authors have come to the conclusion that the effects of DM on the overall reionization process
are relatively small and they cannot compete with those induced by more conventional ionizing sources as stars and 
(mini-) quasars. However, as we will explain in detail later on, radiation from decaying/annihilating DM might change 
the thermal and ionization history of the gas to such an extent that differences with the case in which DM is not 
considered are large enough to produce a clear signature in observable 21~cm signal. Assessing the amplitude of these
deviations from the standard scenario requires a careful study of complex physical processes, and in particular
of the amount of energy that can be transferred from the decay/annihilation products to the IGM.  
A recent study (Ripamonti, Mapelli, \& Ferrara 2006a, hereafter RMF06a) finds that only a relatively small fraction 
of the injected energy is effectively absorbed by the IGM and goes into heating and reionization. As a result, the 
effects on the 21~cm background signal might be smaller than previously predicted (Shchekinov \& Vasiliev 2006; 
Furlanetto et al. 2006); hence, the question remains if the DM signal can still be observed by next generation 
radio interferometers. If so, this experiment would constitute a superb tool to distinguish among 
different DM candidates via their decaying/annihilating properties.  
In this paper we calculate the effects of DM decays/annihilations on the 21~cm background for some of the
most popular DM candidates, such as decaying or annihilating Light Dark Matter (LDM) and decaying sterile neutrinos. 

The rest of the paper is organized as follows. In Sec. 2 we briefly introduce the DM candidates and their effect
on the IGM; in Sec. 3 we provide the basic equations to study the 21~cm radiation in presence of decaying/annihilating 
DM. In Sec. 4 we present the results of our calculations, which are then discussed in Sec. 5.
Throughout the paper we assume, in agreement with the 3-yr WMAP data analysis (Spergel et al. 2006), 
a $\Lambda$CDM cosmology with $\Omega_m$ = 0.24, $\Omega_{\Lambda}$ = 0.76, $h$ = 0.73, $\Omega_b$ = 0.042,
$H_0$ = $100 \, h \, \mbox{km} \, \mbox{s}^{-1} \, \mbox{Mpc}^{-1}$.

\section {Basic physics}

We are interested in calculating the effects on HI 21~cm line signal produced by two among the most popular 
low-mass DM candidates, \ie LDM ($\sim{}1-10$ MeV) and sterile neutrinos ($\sim{}2-50$ keV). In the case 
of sterile neutrinos only the decay process is allowed; LDM particles instead can both decay and annihilate.
We neglect heavier DM candidates (with mass larger than $\sim{}100$ MeV) because previous studies (Mapelli, Ferrara \& 
Pierpaoli 2006) have already shown that, even assuming that all the energy released following annihilations
is immediately absorbed by the IGM, they represent a negligible heating/ionization source for the gas.

Both in the case of sterile neutrinos and of LDM, the rate of energy transfer per baryon to the IGM can be written 
as (RMF06a; Ripamonti, Mapelli \& Ferrara 2006b, hereafter RMF06b):
\begin{equation}\label{eq:energyabs}
\dot{E}_x(z)=f_{abs}(z)\,{}\dot{n}_{\rm DM}(z)\,{}m_{\rm DM}\,{}c^2,
\end{equation}
where $m_{\rm DM}$ is the mass of the DM particle and $c$ is the speed of light. The energy absorbed fraction, 
$f_{abs}(z)$, is discussed in detail in the following;  $\dot{n}_{{\rm DM}}(z)$ is the decrease rate of the number of DM
particles per baryon.  

In the case of DM decays, $\dot{n}_{{\rm DM}}(z)$ is given by
\begin{equation}\label{DM density decay}
\dot{n}_{\rm DM}(z) \simeq {{n_{{\rm
	DM},0}}\over{\tau_{\rm DM}}},
\end{equation}
where $n_{{\rm DM},0}$ and $\tau_{\rm DM}$ are the current number of DM
particles per baryon and the lifetime of DM particles,
respectively.

For the annihilations: 
\begin{equation}\label{DM density annih}
\dot{n}_{\rm DM}(z) \simeq {1\over2}\, n_{{\rm DM},0}^2\,{}{\mathcal N}_{\rm b}(0)\,{}
\langle \sigma\,v\rangle (1+z)^3,
\end{equation}
where ${\mathcal N}_{\rm b}(0)=2.5\times{}10^{-7}\;{\rm cm^{-3}}$ is the current
baryon number density (Spergel et al. 2006), and $\langle\sigma v\rangle$ is the
thermally-averaged DM annihilation cross section. 
The values of $m_{\rm DM}$, $n_{{\rm DM},0}$,  $\tau_{\rm DM}$ and  $\langle\sigma v\rangle$ depend from the 
properties of the chosen DM candidate (see Secs. 2.1 and 2.3).

\subsection{The energy absorbed fraction}
The most delicate term in eq.~(\ref{eq:energyabs}) is represented by $f_{abs}(z)$, \ie the fraction of the DM particle rest mass that is absorbed by the gas at a given redshift $z$; $f_{abs}$ strongly depends on how the decay/annihilation products interact with the IGM. Since accounting for the physical processes which govern such interactions is quite complicated, 
most of previous studies assume that: (i) all the energy released by DM decays/annihilations is immediately absorbed 
(Hansen \& Haiman 2004; Pierpaoli 2004; Biermann \& Kusenko 2006; Mapelli et al. 2006), (ii) leave $f_{abs}$ as a free parameter (Padmanabhan \& Finkbeiner 2005; Zhang et al. 2006), or (iii) make a partial treatment of the energy redistribution 
(Chen \& Kamionkowski 2004; Mapelli \& Ferrara 2005).

Recently, RMF06a have calculated behaviour of  $f_{abs}(z)$ in detail, for the most common case in which the 
decay/annihilation products are photons, active neutrinos or electron-positron pairs. Photons are affected 
by Compton scattering and photo-ionization; for pairs, the relevant processes are inverse Compton scattering,
collisional ionizations, and positron annihilations. 

RMF06a found that, if the decay/annihilation products are either photons and active neutrinos or pairs, $f_{abs}$ is 
close to the maximum allowed value (equal to 0.5 for sterile neutrinos, due to the active neutrino production, 
and equal to 1 for LDM) only at very high 
redshift ($z\gg{}100$). At lower redshifts, $f_{abs}$ rapidly drops to values $<0.1$; also, the higher is the mass of 
the progenitor DM particle, the faster is the decrease of $f_{abs}$.  Thus, accounting for the correct  $f_{abs}(z)$ 
determination dramatically reduces the possible effects of DM decays and annihilations on the IGM heating and ionization 
(RMF06a) with respect to previous estimates. 

In this paper, for the first time, we will adopt the correct estimate of $f_{abs}(z)$ (as given in RMF06a) in order to 
evaluate the impact of DM decays and annihilations on 21~cm emission. Previous papers (e.g. Furlanetto et al. 2006) 
have assumed a redshift-independent $f_{abs}(z)$, which appears to be quite unrealistic (for a discussion, see Sec. 4) 
and leads to optimistic upper limits for the contribution of DM-related effects to 21~cm maps. 

Differently from Furlanetto et al. 2006, who do not select any specific DM candidate and leave the decay rate as a 
free parameter, we chose to consider three specific DM candidates (sterile neutrinos, decaying LDM and annihilating LDM). 
This choice allows us to give predictions which can be more easily related to other DM measurements, such as the X-ray 
constraints on the sterile neutrino mass (Watson et al. 2006; Boyarsky et al. 2006a) or the detection of the 511-keV 
emission line from the Galactic center (Kn\"odlseder et al. 2005).

\subsection{Sterile neutrinos}
Many models of sterile neutrinos have been proposed, with mass ranging from eV to TeV. Here, we are interested in 
sterile neutrinos as warm DM (WDM) candidates, with masses of the order of a few keV ($\sim{}2-50$ keV). These particles 
can decay into an active neutrino and a photon (Dolgov 2002 and references therein).

The mass (and thus the lifetime) of radiatively decaying sterile neutrinos can be constrained by the absence of any detection of X-ray lines 
consistent with photons due to sterile neutrino decays in galaxy clusters (Abazajian, Fuller \& Tucker 2001; 
Abazajian 2006; Abazajian \& Koushiappas 2006; Boyarsky et al. 2006b) or in galaxies (Watson et al. 2006 obtained 
the strongest constraints from the study of the Andromeda galaxy). Other constraints come from the comparison between 
the unresolved X-ray background and the expected contribution from sterile neutrino decays (Mapelli \& Ferrara 2005; 
Boyarsky et al. 2006a).

In this paper we consider the representative case of $m_\nu = 25$~keV sterile neutrinos, whose contribution to heating 
is maximum (RMF06a), due to the weakness of the available constraints on lifetime for neutrinos of such mass. For 
such particle the upper limits on the lifetime and the present number density number per baryon are  
$\tau{}_{\rm DM}=9.67\times{}10^{25}\textrm{ s }$ and $n_{{\rm DM},0}=1.88\times{}10^5$, respectively (see RMF06a, RMF06b). 
The contribution of other sterile neutrinos masses to the 21 cm line is expected to be comparable or smaller 
than for the case considered here.

\subsection{Light dark matter}
We define as LDM particles all the DM candidates whose mass is $1 \le m_{LDM}/{\rm MeV} \le 100$.
Such particles have been suggested as a possible source for the detected 511-keV excess from the Galactic centre 
(Kn\"odlseder et al. 2005). According to this scenario, their maximum allowed mass $m_{\rm LDM}$ should be
20 MeV, not to overproduce detectable gamma rays via internal bremsstrahlung (Beacom, Bell \& Bertone 2004), or 
even $\sim{}$3 MeV, if we consider also the production of gamma rays for inflight annihilations of the positrons 
(Beacom \& Y\"uksel 2006).

LDM can  decay or annihilate, producing photons, neutrinos and pairs. We will treat both channels in detail 
and assume that the only decay/annihilation products are pairs, an assumption leading to an upper limit in terms of 
IGM heating (RMF06a).

We consider, as a template, the case of 10-MeV LDM particles, which again yields the most efficient heating 
case (see RMF06a). For such particles the upper limits of the current number (per baryon), the lifetime and 
the cross-section are $n_{{\rm DM},0}\sim{}446$, $\tau{}_{\rm DM}=4\times{}10^{25}\textrm{ s}$, and 
$\langle{}\sigma{}v\rangle{}\sim{}2.4\times{}10^{-28}\textrm{ cm}^{3}\textrm{ s}^{-1}$, respectively (RMF06a).

\section{Effects on the 21~cm radiation}

\subsection{Ly$\alpha$ pumping}

The 21~cm line is associated with the hyperfine transition between the triplet and the
singlet levels of the hydrogen ground state.
This transition is governed by the spin temperature, $T_{S}$, defined as:
\begin{equation}
\frac{n_{1}}{n_{0}}=3 \exp \left(-\frac{{T_{\star}}}{T_{S}}\right),
\end{equation}
where $n_{0}$ and $n_{1}$ are the number densities of hydrogen atoms in the
singlet and triplet ground hyperfine levels, and $T_{\star}=0.068$~K is the 
temperature corresponding to the transition energy.

In the presence of the Cosmic Microwave Background (CMB) alone, the spin temperature reaches thermal 
equilibrium with $T_{\rm CMB}=2.73(1+z)$~K on a short time-scale, making the HI undetectable
in emission or absorption. 

Two mechanisms can decouple $T_{S}$ from $T_{\rm CMB}$: (i) collisions, which are effective mainly 
at $z\geq $ 70 due to the higher mean IGM density, and (ii) scattering by Ly$\alpha$ photons $-$
the so-called Wouthuysen-Field process or 
Ly$\alpha$ pumping (e.g. Wouthuysen 1952; Field 1959; Hirata 2005) $-$ which couples $T_{S}$ to the 
kinetic gas temperature $T_{K}$ via the  
mixing of the hyperfine levels of the HI ground state through intermediate
transitions to the excited $2p$ state. 

The spin temperature is then given by the equation:

\begin{equation}\label{eq:ts}
T_S=\frac{T_{CMB}+y_\alpha T_k+ y_c T_k}{1+y_\alpha+y_c}
\end{equation}
where $T_{k}$ is the kinetic temperature, and the Ly$\alpha$ and collisional coupling coefficients are given by the
following expressions:

\begin{equation}
y_\alpha=\frac{P_{10}T_\ast}{A_{10} T_k},
\end{equation}

\begin{equation}
y_c=\frac{C_{10}T_\ast}{A_{10} T_k},
\end{equation}
where

\begin{equation}\label{eq:c10}
C_{10}=k_{10}n_{HI}+n_e {\gamma}_e,
\end{equation}
and

\begin{equation}
P_{10}=\frac{16}{27} \frac{\pi J_\alpha {\sigma}_\alpha}{h_p {\nu}_\alpha}.
\end{equation}
In the above equations
$A_{10}=2.85 \times 10^{-15} {\rm s}^{-1}$ is the spontaneous emission coefficient of the 21~cm line, 
$P_{10}$ is the indirect de-excitation rate of the
hyperfine structure levels. We write $P_{10}=4 P_\alpha/27$; in addition, 
$P_\alpha=(4 \pi J_\alpha {\sigma}_\alpha)/(h_p {\nu}_\alpha)$ is the outcome of the
equation for the Ly$\alpha$ scattering rate,
\begin{equation}
P_\alpha= c \int{d\nu \,n(\nu) \frac{\sigma(\nu)}{\nu}}=\frac{4 \pi}{h} \int{d\nu \, \frac{J_\nu \, \sigma(\nu)}{\nu}},
\end{equation}
in the case that $\sigma(\nu)$ is a $\delta$ function. A detailed investigation of the 
physics of the Wouthuysen-Field process (Hirata 2005) finds small corrections to these expressions 
which we neglect here for simplicity.  
The coefficient $C_{10}$ in eq.~(\ref{eq:c10}) is the collisional de-excitation rate by hydrogen atoms and electrons, 
where $k_{10}$ is tabulated in Allison \& Dalgarno (1969) for different temperatures; 
$\gamma_e$, according to Liszt (2001), is 
\begin{equation}
\log{\gamma_e(T_k)}=-9.607+ \log{T_k}^{1/2} \exp(-(\log T_k)^{4.5}/1800).
\end{equation}
for $T_k \leq 10^4$~K, while $\gamma_e(T_k > 10^4 \mbox{K})=\gamma_e(T_k=10^4 \mbox{K})$. 

\begin{figure*}
  \centerline{\psfig{figure=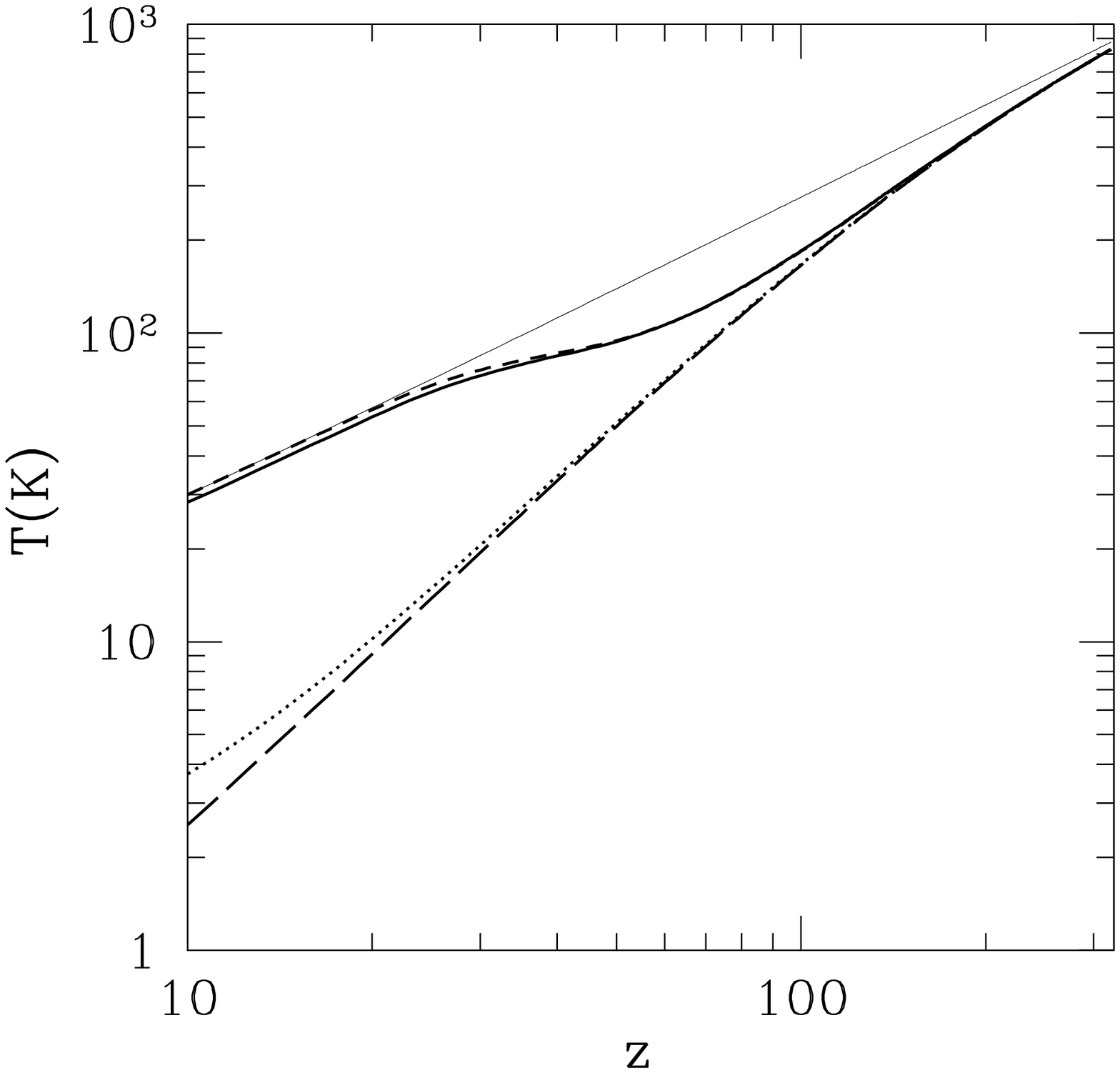,width=8.5cm,angle=0}
    \psfig{figure=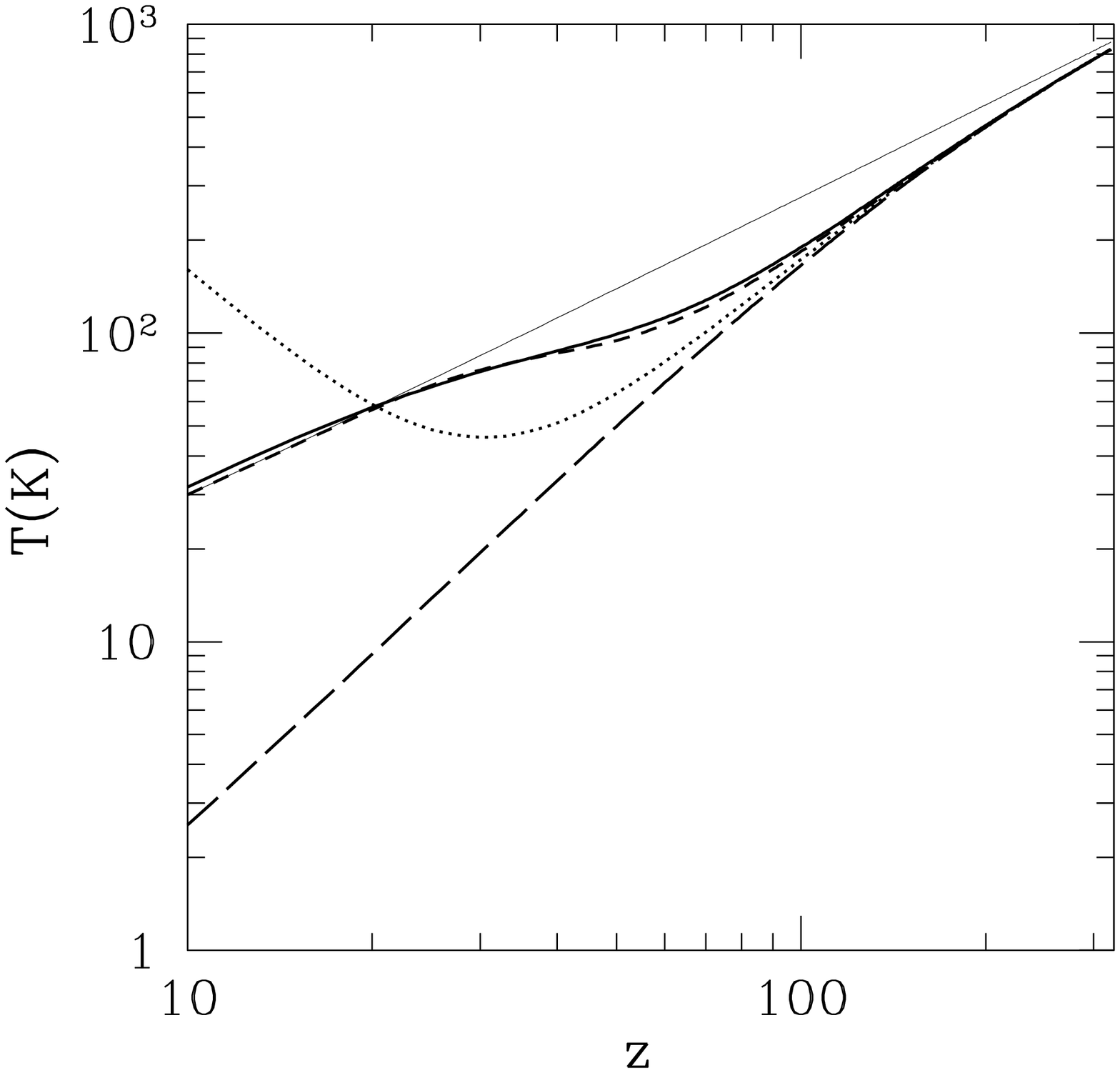,width=8.5cm,angle=0}}
  \caption{{\it Left panel}: $T_{S}$ and $T_k$ as a function of redshift (25-keV sterile neutrino decays, $f_{abs}$ as in RMF06a). 
Thin solid line: $T_{CMB}$; 
thick solid (short-dashed) line: $T_{S}$ with (without) 25-keV neutrino decays;
dotted (long-dashed) line: $T_k$ with (without) 25-keV neutrino decays. {\it Right}: $T_{S}$ and $T_k$ as a function of redshift for 25-keV sterile neutrino decays 
and for $f_{abs}=0.5$. The lines are the same as in the left panel.}
\label{fig:1}
\end{figure*} 

For the Ly$\alpha$ pumping to be effective, a minimum Ly$\alpha$ background intensity, $J_{\alpha}$, 
is required. This is given by the condition (Ciardi \& Madau 2003): 
\begin{equation}
J_{\alpha}\geq 9 \times 10^{-23}(1+z)\,\mbox{erg}\,\,\mbox{cm}^{-2}\,\mbox{s}^{-1}\,
\mbox{Hz}^{-1}\, \mbox{sr}^{-1}, 
\end{equation}
at the redshift of interest.

\subsection{IGM evolution and 21~cm background.}

Next, we want to understand how DM decays/annihilations affect the thermal and ionization evolution of the 
IGM, which, in turn, determines the level of the 21~cm background signal.
The equation that describes the evolution of the ionized fraction $x_{e}$ is the following (see e.g.
Chen \& Kamionkowski 2004):   
\begin{eqnarray}
-\frac{dx_{e}}{dz} & = & \frac{1}{H(z)(1+z)}[R_{s}(z)-I_{s}(z)-I_{x}(z)],
\end{eqnarray}
where $I_{x}={\dot{E}}_{x}/E_{0}$ is the contribution to the ionization rate due to DM; 
$I_{s}$ and $R_{s}$ are the standard ionization and recombination rates per baryon.
Considering eqs. (1)-(3) we have that
\begin{eqnarray}
I_{x}=\chi_{i}(z) \frac{{\dot{E}}_{x}}{E_{0}}=f_{abs}(z) \chi_{i}(z)\Gamma_{x} f_{x} \frac{m_p c^2}{E_{0}} 
\end{eqnarray}
for DM decays, and
\begin{eqnarray}
I_{x}=\chi_{i}(z) \frac{{\dot{E}}_{x}}{E_{0}}=f_{abs}(z) \chi_{i}(z) f_{x} \frac{m_{p} c^2}{E_{0}} n_{DM,0} {\mathcal N}_{b}(z) \langle \sigma\,v\rangle
\end{eqnarray}
for DM annihilations.
In the last two equations $E_{0}=13.6$~eV is the hydrogen ionization threshold, $m_p$ is the proton mass, $\chi_{i}$ is the fraction of the 
energy absorbed by the IGM from DM decays/annihilations that goes into ionizations ($\chi_{i}\sim (1-x_{e})/3$, see
Shull 1979, Shull \& van Steenberg 1985), $f_{x}=\Omega_{x}(z)/\Omega_{b}(z)$, ${\mathcal N}_{b}(z)$ is the baryon number density and 
$\Gamma_{x}=1/\tau_{\rm DM}$.

The equation regulating the evolution of IGM temperature can be written as:
\begin{eqnarray}
(1+z)\frac{dT_{k}}{dz}&=&2T_{k}+\frac{l_{\gamma } x_{e}}{H(z)(1+f_{He}+x_{e})}(T_{k}-T_{CMB}) \nonumber\\
                   &   & \nonumber \\
                   &   & - \frac{2 \chi_h \dot E_x}{3k_{b} H(z) (1+f_{He}+x_{e})}
\end{eqnarray}\newline
where $l_{\gamma }=(8\sigma_{T} a_{R} T_{CMB}^4)/(3 m_{e} c)$, $\chi_{h}$ is the fraction of the absorbed DM energy
deposited into the IGM as heating (Shull \& van Steenberg 1985) and $f_{He}=$ 0.24 is the helium fraction by mass. The Ly$\alpha$ heating 
resulting from repeated scatterings as the photons are redshifted into the 
Lyman resonances is negligible as it has been recently shown (see e.g. Chen \& Miralda Escud$\acute{e}$ 2004).
The ionization and temperature eqs. (13)-(16) are solved using a modified version of RECFAST (Seager et al. 1999).

A third equation is needed in order to compute the 21~cm backgroundi: the one describing the evolution of
the Ly$\alpha$ background intensity $J_\alpha$. Following Madau, Meiksin \& Rees (1997) we write:
\begin{equation}
J_{\alpha}(z)=\frac{{\mathcal N}_{H}^2hc}{4\pi H(z)}\left[x_{e}x_{p}\alpha_{{2^2P}}^{eff}+x_e x_{HI} \gamma_{eH}+\frac{\chi_{\alpha} \dot{E}_x(z) }{{\mathcal N}_H h \nu_{\alpha}}\right],
\end{equation}
where the first two terms are the contributions from recombinations and collisional excitations by electron impacts,
while the third term is the DM contribution.

\begin{figure*}
  \centerline{\psfig{figure=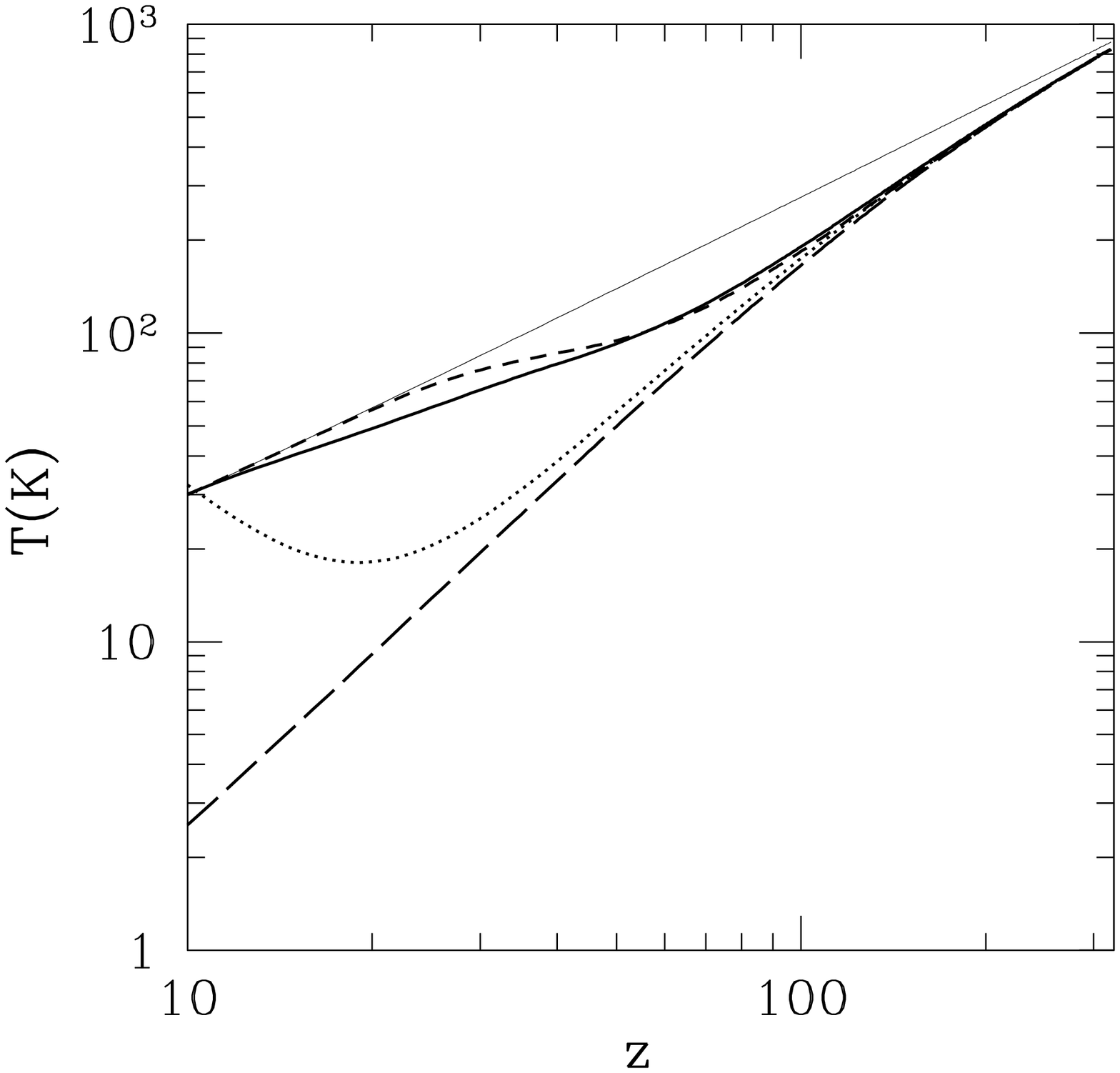,width=8.5cm,angle=0}
    \psfig{figure=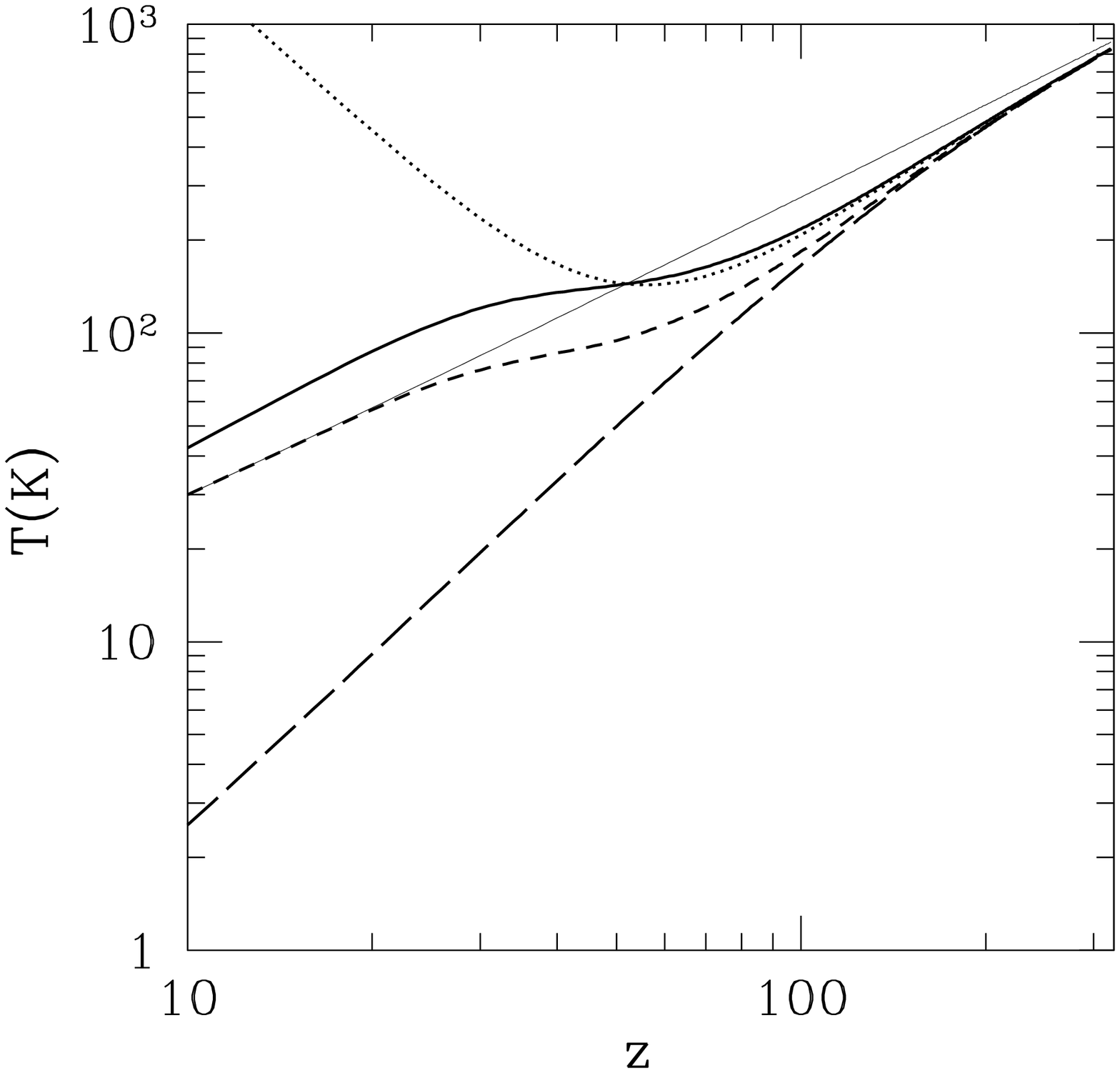,width=8.5cm,angle=0}}
  \caption{{\it Left panel}: $T_{S}$ and $T_k$ as a function of redshift (10-MeV LDM decays, $f_{abs}$ as in RMF06a). 
Thin solid line: $T_{CMB}$; 
thick solid (short-dashed) line: $T_{S}$ with (without) 10-MeV LDM decays;
dotted (long-dashed) line: $T_k$ with (without) 10-MeV LDM decays. {\it Right}: $T_{S}$ and $T_k$ as a function of redshift for 10-MeV LDM decays 
and for $f_{abs}=1$. The lines are the same as in the left panel.}
\label{fig:2}
\end{figure*}

In the last equation ${\mathcal N}_{H}$=$0.92 {\mathcal N}_{b}$ is the number density of hydrogen atoms, 
$\alpha_{{2^2P}}^{eff}$ 
is the effective recombination coefficient to the $2^2P$ level (Pengelly 1964), which
includes direct recombinations to the $2^2P$ level and recombinations to higher levels, 
followed by transitions to the $2^2P$ level via all possible cascade paths. 

Finally, $\gamma_{eH} \approx 2.2 \times 10^{-8} \exp{(-11.84/T_4)} \, \mbox{cm}^3 \mbox{s}^{-1}$ 
is the collisional excitation rate of HI atoms photons by electron impacts; 
$\chi_\alpha$ is the net fraction of the absorbed X-ray photons from DM decays/annihilations
which is converted into Ly$\alpha$ photons (Shull \& van Steenberg 1985), and $T_4=T/(10^4\textrm{ K})$.

Once $T_S(z)$ has been determined through eq.~(\ref{eq:ts}), we can obtain the
21~cm radiation intensity, which can be expressed by the differential brightness temperature
between a neutral hydrogen patch and the CMB:
\begin{equation}
\delta T_{b}\,\simeq\,\frac{T_{S}-T_{CMB}}{1+z}\,\tau,
\end{equation}
where $\tau$ is the optical depth of the neutral IGM at $21(1+z)$~cm:
\begin{equation}\label{eq:tau}
\tau\,\simeq \,\frac{3c^{3} h_{p} A_{10}}{32 \pi k_{B} {\nu_{0}}^{2} T_{S}H(z)}{\mathcal N_{\rm HI}}.
\end{equation}
In equation~(\ref{eq:tau}), $h_{p}$ and $k_{B}$ are the Planck and Boltzmann constants, respectively, $\nu_{0} = 1420$~MHz
is the 21~cm hyperfine transition frequency, and ${\mathcal N}_{\rm HI}$ is the local HI number density.
If $T_{S}$ is higher than $T_{\rm CMB}$, the neutral IGM will be visible in emission 
against the CMB; on the contrary, if $T_{S} < T_{\rm CMB}$ it will be visible in absorption.

\section{Results}
In a Universe where DM does not decay or annihilate the spin temperature and the kinetic temperature of the gas 
track the CMB temperature down to $z\approx $ 300, when the kinetic temperature starts to decrease adiabatically, 
$T_{K}\propto (1+z)^{2}$, while $T_{CMB}\propto (1+z)$. The collisions at this redshift are efficient at coupling 
$T_{S}$ and $T_{K}$: the spin temperature subsequently tracks the kinetic temperature down to $z\sim $ 70. 
At lower redshifts radiative coupling to the CMB becomes dominant and $T_{S} \rightarrow T_{CMB}$ again.
As a result, for 30 $\leq z\leq $ 300 HI is visible in absorption against the CMB at wavelength of 21 $(1+z)$ cm.
This scenario could change considerably if we allow for DM decay/annihilation, depressing or even erasing the
absorption feature discussed above. Also, we note that different DM candidates leave different traces on 
the 21~cm background signal, and therefore it could be possible, in principle, to constrain directly the DM 
nature through 21~cm observations. To isolate the effect of DM, we will assume in the following that the entire
cosmic dark matter content is constituted by particles of the considered type (sterile neutrinos or LDM), 
\ie $\Omega_X=\Omega_m-\Omega_b$.

\subsection{WDM: 25-keV sterile neutrino decay}

In Fig. 1 we compare the effects of 25-keV sterile neutrino decays on $T_{S}$ assuming either the 
physically motivated value of $f_{abs}$ (left panel) or the commonly adopted value $f_{abs}=0.5$ (right panel), 
i.e. the maximum allowed value for sterile neutrinos corresponding to complete absorption. 
In addition, in each of the panels we explore the differences between models with or without the 
energy injection term due to DM decays. We will refer in the following to the latter as the {\it standard} case,
for brevity.

With respect to the case without DM energy injection, we find, independently of the assumption
made for $f_{abs}$, a higher kinetic temperature and hydrogen ionization 
fraction. However, such an enhancement produces only very modest $T_S$ differences with respect 
to the standard case, which has some effect on the differential brightness temperature $\delta T_{b}$. 
The difference $T_S-T_{CMB}$ remains extremely small due to the limited ability of
the Ly$\alpha$ pumping to decouple the two temperatures.

\begin{figure*}
  \centerline{\psfig{figure=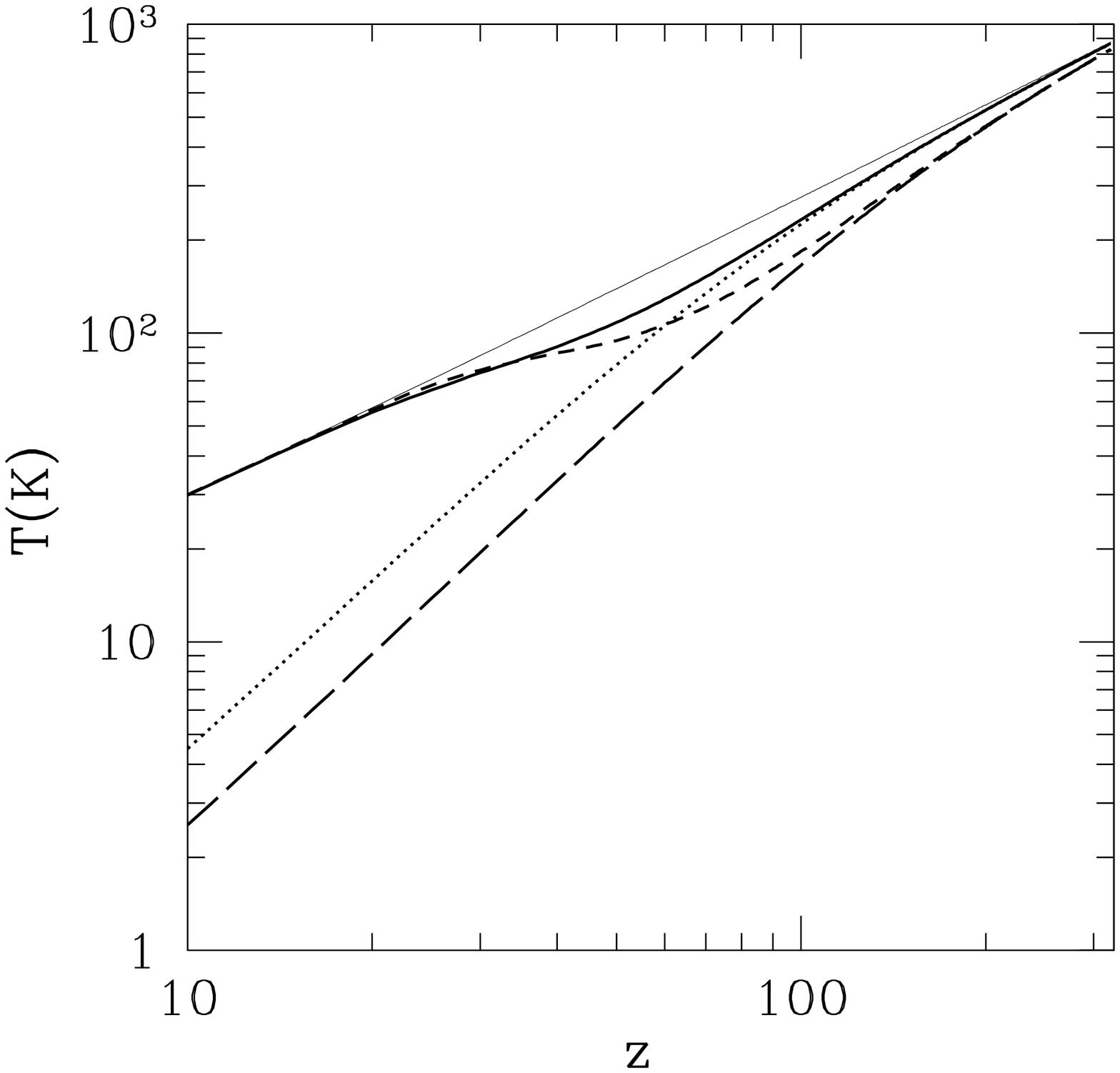,width=8.5cm,angle=0}
    \psfig{figure=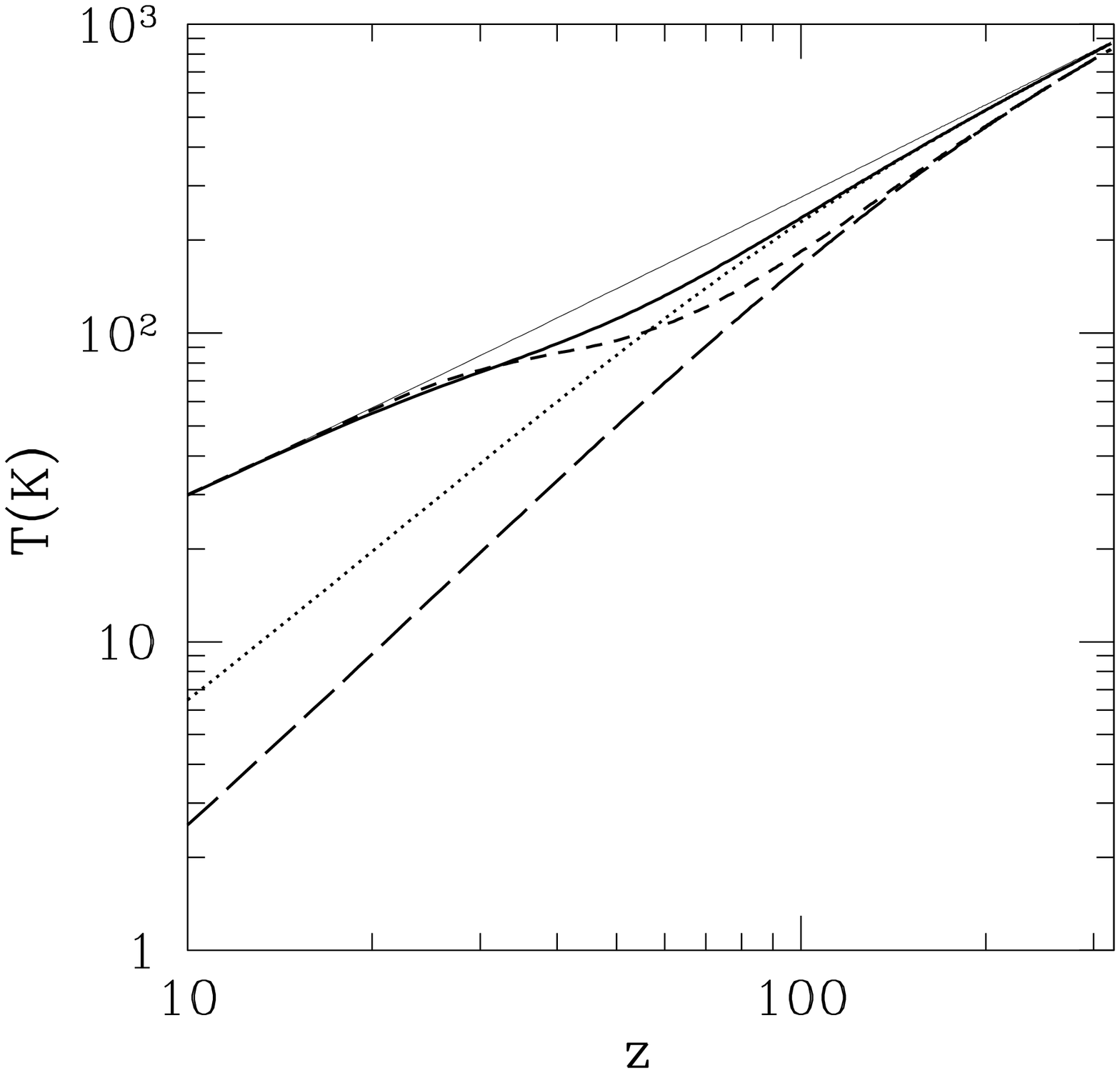,width=8.5cm,angle=0}}
  \caption{{\it Left panel}: $T_{S}$ and $T_k$ as a function of redshift (10-MeV LDM annihilations, $f_{abs}$ as in RMF06a). 
Thin solid line: $T_{CMB}$; 
thick solid (short-dashed) line: $T_{S}$ with (without) 10-MeV LDM annihilations; dotted (long-dashed) line: 
$T_k$ with (without) LDM annihilations. {\it Right}: $T_{S}$ and $T_k$ as a function of redshift for 10-MeV 
LDM annihilations and for $f_{abs}=1$. The lines are the same as in the left panel.}
\label{fig:3}
\end{figure*}

The main difference between the results obtained from the two $f_{abs}$ assumptions   
is the behavior of $T_{K}$ at $z\sim $ 10$-$70. When we use the realistic $f_{abs}$value, 
the IGM kinetic temperature positively deviates at most by 50\% from its standard evolution; 
if instead $f_{abs}=0.5$ is adopted (e.g. as assumed for example by Furlanetto et al. 2006), 
the kinetic temperature overshoots the CMB one and reaches very large values ($> 100$~K) at
low redshifts. 

\subsection{LDM: 10-MeV decay/annihilation}

We now turn to the analysis of LDM candidates. Following the same procedure as in the previous Section we
compare the results of the two different assumptions for $f_{abs}$ (note that for LDM, the constant absorbed fraction 
case corresponds to $f_{abs}=1$, see Sec. 2.1), and also test the effects of decays/annihilation 
against the standard case. 

We start from the analysis of the realistic $f_{abs}$ case (Fig. 2, left panel) for decaying LDM. The most striking feature 
is that the evolution of $T_K$ decouples from the adiabatic one already at $z\approx 50$, and starts
to increase below $z\approx 20$, reaching $T_K=30$~K at $z=10$. Such thermal history forces $T_S$
to remain below $T_{CMB}$ for a much longer redshift interval (down to $z=10$) than in the standard case, 
in which $T_S \approx T_{CMB}$ already at $z\approx 25$. Interestingly, this effect extends the frequency
range in which the IGM can be observed in absorption to higher values.  The main physical reason for the
larger impact of LDM decays on $T_S$ with respect to sterile neutrinos analyzed above basically lies in
the larger $f_{abs}$ value (approximately a factor of 10 below $z=30$) of LDM particles; this is
on top of their larger rest mass.
The higher heating rate increases both $T_K$ and the Ly$\alpha$ 
background (\ie $y_\alpha$), thus pushing $T_S$ away from $T_{CMB}$.   

When $f_{abs}$ is instead artificially forced to be constant and equal to unity (Fig. 2, right panel), 
the physical arguments given above can still be used to interpret the results, but the 
deviation from the standard case is much more dramatic, and results in an overestimate of the
LDM decays impact.
For example, $T_K$ becomes larger than $T_{CMB}$ already at $z\approx 50$
and stays above it thereafter, reaching $>1000$ K at $z=10$. The increase is so strong that it 
essentially erases the absorption feature expected in the standard case above $z=30$. 
The IGM according to this prescription should be observed in emission up to 
very high redshifts.

For comparison sake, we note that this case has nearly the same decay rate, $\Gamma{}_X=2.5\times{}10^{-26}\textrm{ s}^{-1}$, 
as that assumed in one of the cases explored by Furlanetto et al. (2006) (dot-dashed line in Fig. 1 of that paper). 
Consistently, the evolution of $T_K$ from the two studies is essentially the same. 

Finally we turn to the case of annihilating LDM (Fig. 3). As usual, the left panel of that figure reports
the case in which the realistic, redshift dependent, values of $f_{abs}$ are assumed.  
In this case, the kinetic temperature deviates from the standard adiabatic evolution at extremly
high redshifts ($z> 200$): this is because the annihilation process depends on the square of the
baryon density. As a result, its contribution predominantly occurs at early epochs and progressively
vanishes with time, as realized also from the fact that below $z=100$ the $T_K$ curve is simply shifted
to an higher adiabat. This fact causes the spin temperature to remain closer to (although always lower than)
$T_{CMB}$, thus preserving the standard absorption feature. 
Quite remarkably the results assuming the realistic $f_{abs}$ values do not differ appreciably from 
those obtained by imposing $f_{abs}=1$ (Fig. 3, right panel). As pointed out before, the effects of
annihilating DM are evident only at very high redshift, where the detailed calculation gives $f_{abs} 
\approx 1$ (see RFM06a), thus making the non-evolving $f_{abs}$ approximation acceptable.     

\begin{figure*}
  \centerline{\psfig{figure=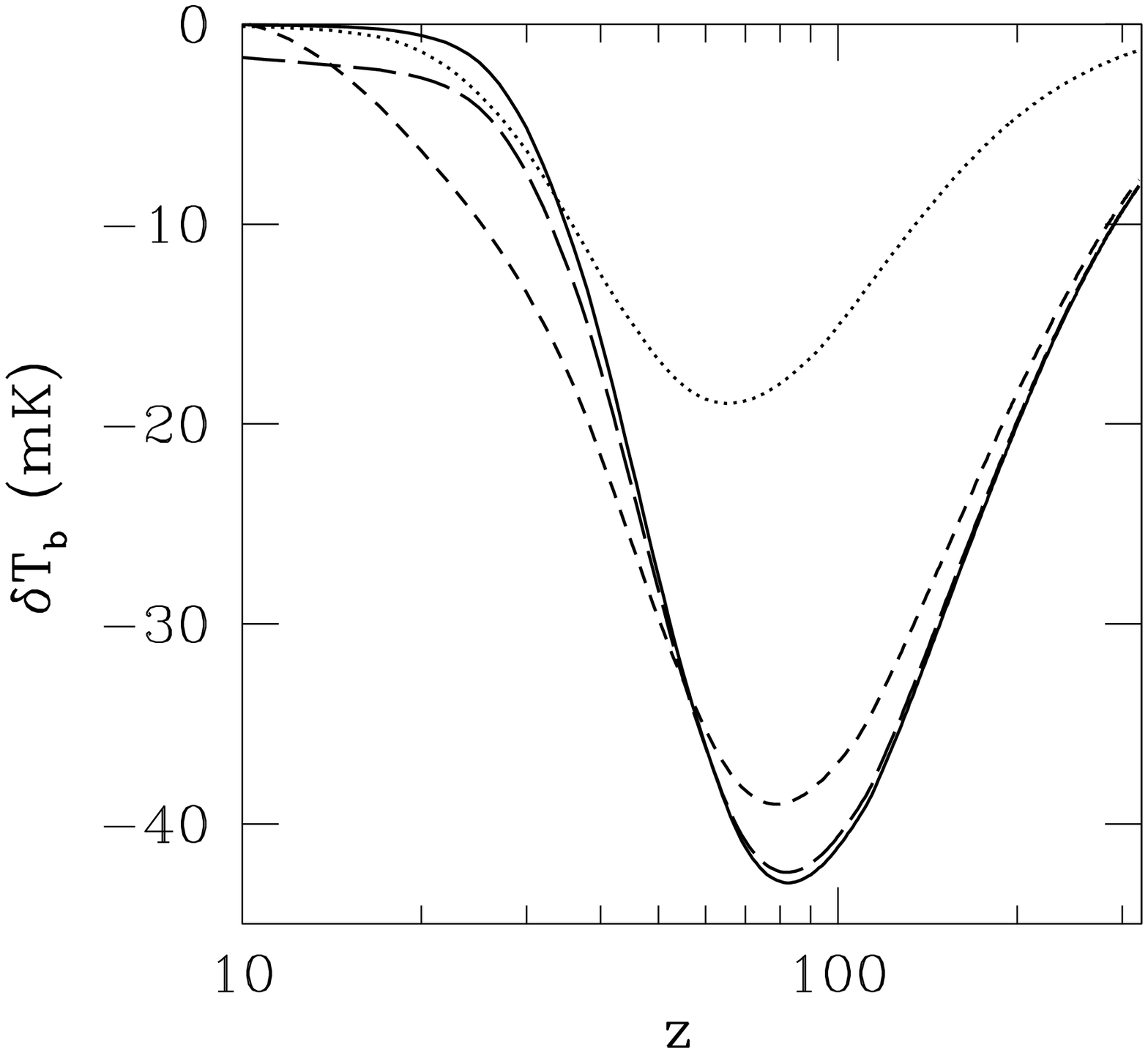,width=8.5cm,angle=0}
    \psfig{figure=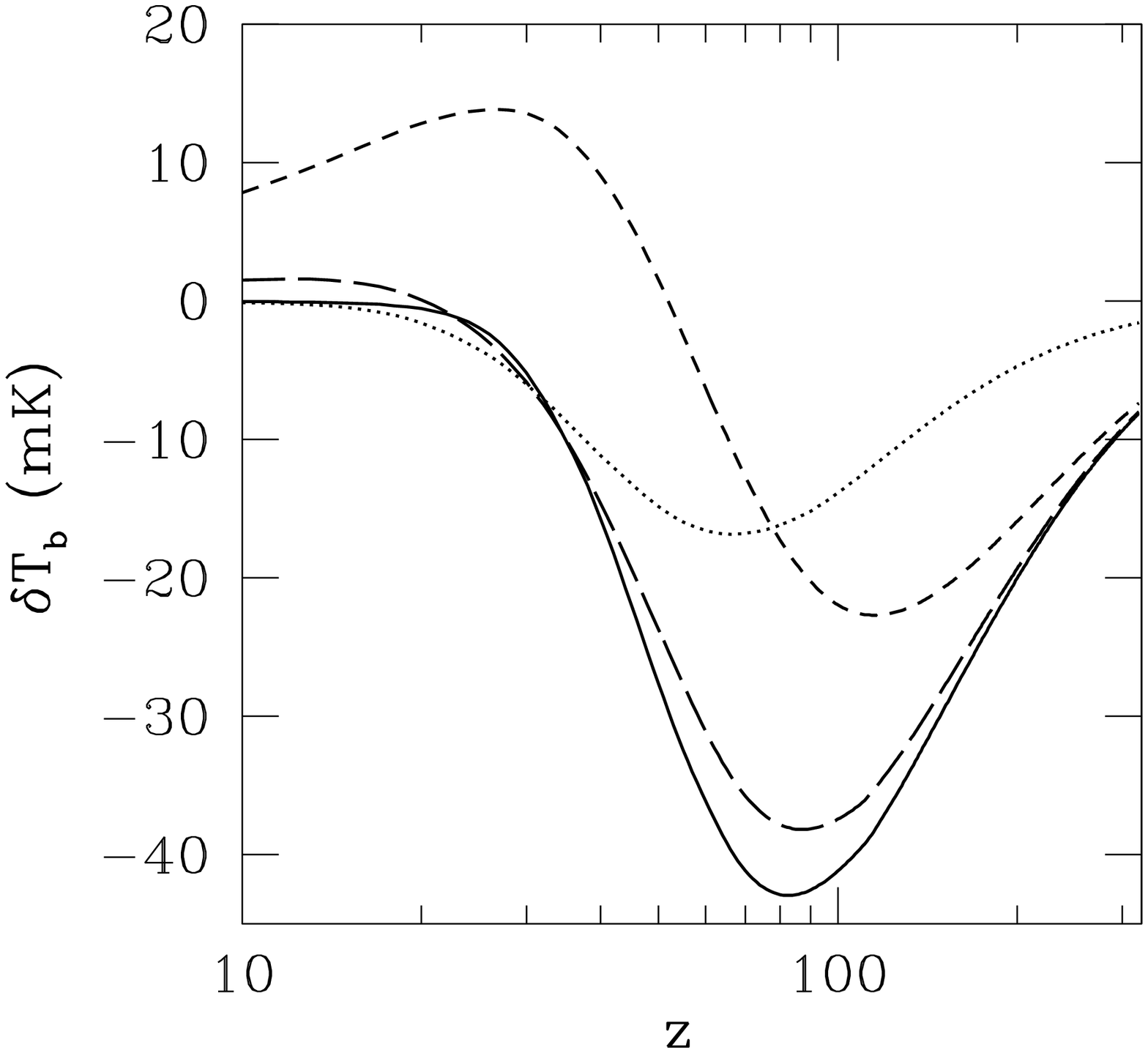,width=8.5cm,angle=0}}
  \caption{{\it Left panel}: 21~cm differential brightness temperature as a function of redshift. The solid line shows $\delta T_{b}$ without 
decaying/annihilating DM; while the long dashed, short dashed and dotted lines refer to $\delta T_{b}$ with  25-keV decaying WDM, 
 10-MeV decaying LDM and 10-MeV annihilating LDM, respectively. Calculations were performed assuming $f_{abs}$ as in RMF06a. 
{\it Right}: 21~cm differential brightness temperature as a function of redshift as in the left panel. 
Calculations were performed assuming $f_{abs}=0.5$ for sterile neutrinos and $f_{abs}=1$ in the other cases.}
\label{fig:4}
\end{figure*}

\subsection{Global 21~cm background.}

Having obtained the evolution of $T_S$ for the different DM candidates, we are now ready to compute
for each case the quantity that is most readily associated with observations, \ie the differential brightness
temperature, $\delta T_{b}$. These are shown in Fig. 4 for the realistic (left panel) 
and the constant $f_{abs}$ (right panel) cases, respectively. In addition, to facilitate the comparison, 
we have also plotted in Fig. 5 the brightness temperature deviation $\delta T_b-\delta T_{b,0}$ of the 
various models from the standard one, restricted to the realistic $f_{abs}$ evolution only. 

As usual, we start our analysis from the realistic $f_{abs}$ case.
For sterile neutrinos (long-dashed curves)
the characteristic absorption feature in the 21~cm background signal expected at 
30 $\leq z\leq $ 300 is only slightly modified from the standard case  
(solid line): the maximum difference is found to be only $\approx -2$~mK in the range $z\sim $10$-$40.
Yet, such a small signature of WDM decays could still be in principle observable: even modest sized 
single-dish radio telescopes can reach the required mK sensitivity in an all sky observation.
The real challenge for the observation is represented by the ability to disentangle this cosmological signal 
from the various foregrounds (particularly the Galactic synchrotron emission) which could be several orders 
of magnitude brighter.  
The case of decaying 10-MeV LDM (short-dashed lines in Fig.4) is more interesting, as the
difference with the standard case are larger. For this case we find  that
the deviation is predicted to be $\approx -(5{\rm -}8)$~mK in the range $20 < z < 40$.
Such an amplitude could be detected by LOFAR or SKA after a 1000 hour all sky integration and 
by most single dish radio telescopes, provide foreground contamination is taken care of. 
Finally, when LDM annihilations are considered, a considerably different $\delta T_b$ evolution
is obtained with respect to the standard case. Larger deviations are present in the entire range
$40 < z < 200$, where the energy input of LDM annihilations forces $\delta T_b$ to values larger 
than $-20$~mK, about two times smaller (in absolute value) than for the standard evolution.    
Such a large difference could in principle facilitate discriminating the annihilation scenario
from the standard one. In practice, though, foregrounds and ionospheric contamination become more 
severe at the low observing frequencies implied by these high redshifts. 

Although of academic interest only, it is instructive to compare the constant $f_{abs}$ cases (Fig. 4, right panel) 
with the previous results. As seen clearly form the Figure, considerably different conclusions would be drawn.
First, sterile neutrino would drive $\delta T_b$ to positive values, resulting in an emission signal below
$z=20$; a similar trend is also obtained for LDM decays, whose effects are seen in 21~cm emission with an
amplitude of about 10 mK. The results for LDM annihilation instead do not show appreciable dependencies on    
the $f_{abs}$ prescription, for the reasons discussed above.

\section{Discussion}

From the results obtained in this paper, we conclude that it is in principle possible to observe
the HI 21~cm signal from the Dark Ages produced by the energy input due to decays/annihilations
of the most popular light/warm DM candidates. If so, radio observations might represent one of the
most promising tools to study the nature of DM, as different particles are predicted to leave
a specific signature on the signal. The sensitivity required to measure the 21~cm background 
signal can be achieved not only by the next generation of radio interferometers, but also by 
existing radio observatories.

However, the various foregrounds (\ie Galactic free-free and synchrotron emission, unresolved
extra-galactic radio sources, free-free emission from ionizing sources, synchrotron emission from 
cluster radio halos and relics) are much stronger than the cosmological signal, 
and will certainly prove extremely difficult to remove. Hence, a clear detection could be challenging 
to achieve.

It is beyond the scope of this paper to deal in detail with the problems of ionospheric scintillation 
and foreground contamination.  A number of studies have discussed the foregrounds complications in some
detail (e.g. Shaver et al. 1999; Oh \& Mack 2003; Di Matteo, Ciardi \& Miniati 2004) and we refer the 
reader to those papers for more information.  For our aims it is sufficient to say that in general
the sky temperature can be roughly described as:
\begin{equation}
T_{sky}\,\sim \, 180 \left(\frac{\nu}{180 \mbox{MHz}}\right)^{-2.6} \,\mbox{K}
\end{equation}
(see e.g. Furlanetto, Oh \& Briggs 2006) and is, at the frequencies of interest, several orders of 
magnitude higher than the cosmological 21~cm background signal.

The success of the 21~cm background observations to constrain DM will depend on the capability
of effectively removing the foregrounds and of correcting for ionospheric variations (e.g. Gnedin \& 
Shaver 2004; Zaldarriaga, Furlanetto \& Hernquist 2004; Santos, Cooray \& Knox 2005; Morales, Bowman \& Hewitt
2005).

An alternative method to study the 21 cm backgrounds is by separating its spectral features from
the smooth foreground spectrum (e.g. Furlanetto 2006, Shaver et al. 1999). 
Fig. 6 shows the gradient of  $\delta T_{b}$ for the DM candidates studied here.
It is clear from the figure that the gradients are higher and easier to constrain in the frequency range
$\nu=1420/(1+z) \mbox{MHz}\sim $ 10$-$40 MHz.
The foreground gradients however increase with decreasing frequency
making it difficult to predict the best frequencies to get to for a successful
detection. 

Discriminating among different DM cases would imply the ability to distinguish the difference 
in brightness temperature and/or in its gradient with respect to the standard scenario.
We define these two quantities as $\Delta \delta T_b=\delta T_b-{\delta T_b}^0$ and 
$\Delta \delta T_b/\Delta  \nu=d \delta T_b/d \nu -{{d \delta T_b}^0/d \nu}$, where 
the $0$ stands for the standard (\ie non decaying/annihilating) DM scenario.

Fig. 7 shows $\Delta \delta T_b$ versus $\Delta \delta T_b/\Delta \nu$ for the three DM models:
the blue triangles correspond to 10-MeV annihilations, the red squares to 10-MeV decays
and the black stars to the 25-keV sterile neutrino decays.
The points are separated by 1 MHz steps in a range that goes from 5 to 130 MHz.
As the sensitivity is somewhat uncertain for the different proposed 21~cm experiments, we
assume as a guideline that a successful detection requires $\Delta \delta T_b>3$~mK and 
$\Delta \delta T_b/\Delta \nu >0.6$~ mK/MHz, respectively. The adopted value for $\Delta \delta T_b$ 
is a conservative one given that mK sensitivities are already achievable in all sky observations even 
by existing single dish radio telescopes. It has been chosen so to partially take into consideration 
the difficulties associated with foregrounds removal.

The shaded area covers the parameter space corresponding to likely non-detections.
For the 25-keV sterile neutrinos case the points lie almost entirely within the shaded area, thus
future observations would hardly be able to pin down this scenario. 

In the case of decaying or annihilating 10-MeV LDM positive detections could be achieved on different
frequency ranges. 
For the decays the ideal frequencies to study would be 40$-$80~MHz, where $\Delta \delta T_b \gsim 4$~mK, 
and 80$-$90~MHz, where $\Delta \delta T_b\sim 2$~mK but $\Delta \delta T_b/\Delta  \nu \gsim 0.6$~mK/MHz.
At $\nu \geq 90$~MHz, or $z\leq 15$ the first luminous sources start having a dominant impact on the 
IGM and the small modifications from decaying/annihilating DM would not be visible.

For the 10-MeV LDM annihilations case the best observing frequencies seem to be 10$-$30~MHz. At these 
frequencies $\Delta \delta T_b \gsim $ 6~mK with a peak of over 20~mK at 10$-$20~MHz. 
Observations at such low frequencies are particularly difficult and probably future 
radio interferometers will not reach such large wavelengths.
Probably 30~MHz is the minimum frequency which will be achieved by next generation instruments, and 
the frequency at which it will then be possible to constrain 10-MeV annihilating LDM.

We conclude then that it is likely that future observations of the 21 cm background radiation
will allow us to effectively constrain among some of the DM candidates, which by decays 
or annihilations can influence deeply the properties of the IGM. 

Nevertheless our study shows that accounting for the physical evolution of $f_{abs}$($z$) strongly reduces
the amplitude of DM effects on the IGM with respect to those predicted by previous studies: for example we find that 
the effect of 25-keV decaying sterile neutrinos would be hardly observable.

\begin{figure}
\begin{center}
\includegraphics[width=8.5cm]{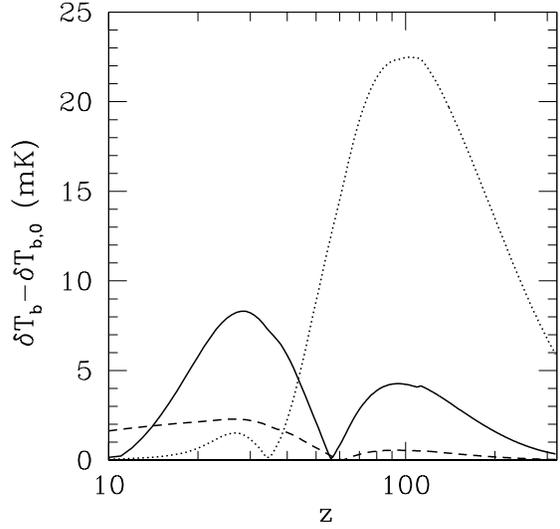}
\end{center}
\caption{21~cm differential brightness temperature difference in mK between the three DM cases considered
and a non decaying$-$annihilating 
DM scenario. The dashed, solid and dotted lines correspond to 25-kev sterile neutrino decays, 
10-Mev LDM decays and 10-MeV annihilations case respectively.}
\label{graph9}
\end{figure}

\begin{figure}
\begin{center}
\includegraphics[width=8.5cm]{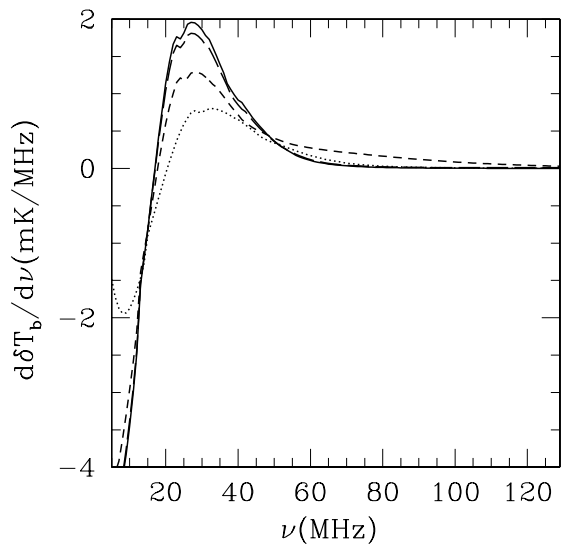}
\end{center}
\caption{Gradient of the 21~cm differential brightness temperature. 
The solid line shows $d \delta T_{b}/d \nu$ without 
decaying/annihilating DM; while the long dashed, short dashed and dotted lines refer to $d \delta T_{b}/d \nu$ 
with  25-keV decaying WDM, 
 10-MeV decaying LDM and 10-MeV annihilating LDM, respectively. Calculations were performed assuming $f_{abs}$ as in RMF06a.}
\label{graph10}
\end{figure}

\begin{figure}
\begin{center}
\includegraphics[width=8.5cm]{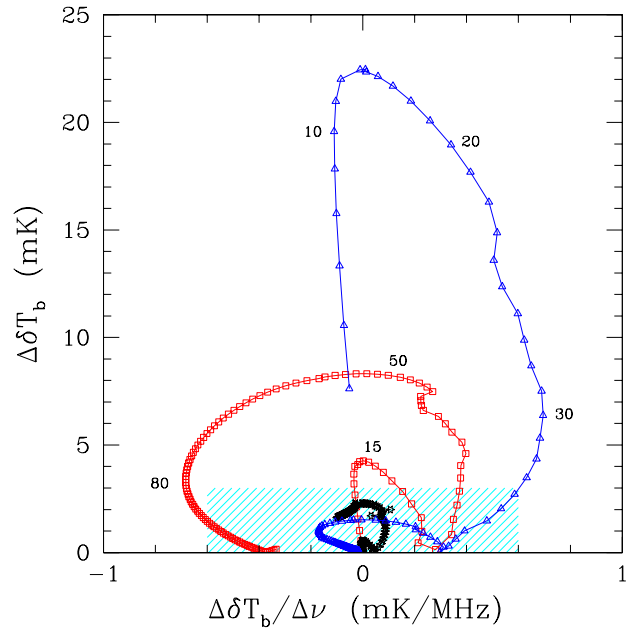}
\end{center}
\caption{Deviations of the differential brightness temperature ($y$-axis) and its gradient ($x$-axis)
from those predicted in the standard case.
The diagram illustrates the capability of constraining DM with future 21~cm observations.
The black stars, red squares and blue triangles represent 25-keV decaying WDM, 
 10-MeV decaying LDM and 10-MeV annihilating LDM, respectively. The shaded area indicates the parameter space in which
detection is excluded given the expected observational capabilities. Numbers along the curves refer to frequencies in 
MHz units.}
\label{graph11}
\end{figure}

%\section*{Acknowledgments}

\end{document}